\documentclass[aps,prl,twocolumn,superscriptaddress,showpacs,amsmath,amssymb,longbibliography]{revtex4-1}
\usepackage{graphicx}
\usepackage[breaklinks]{hyperref}

\usepackage[utf8]{inputenc}
\usepackage{polski}
\usepackage[polish,english]{babel}
\usepackage{amsmath}
\usepackage{xcolor}

\definecolor{nred} {RGB}{224,0,0}
\definecolor{nblue} {RGB}{28,130,185}
\definecolor{dgreen} {RGB}{78,138,21}
\definecolor{norange}{RGB}{230,120,20}

\makeatletter
\newcommand*{\balancecolsandclearpage}{%
  \close@column@grid
  \clearpage
  \twocolumngrid
}

\usepackage[sort&compress]{natbib}
 
\begin{document}

\title{Counting local integrals of motion in disordered spinless--fermion and Hubbard chains}
\author{Marcin Mierzejewski}
\affiliation{Institute of  Physics,  University  of  Silesia,  40-007  Katowice,  Poland}
\author{Maciej Kozarzewski}
\affiliation{Institute of  Physics,  University  of  Silesia,  40-007  Katowice,  Poland}
\author{Peter Prelov{\v{s}}ek}
\affiliation{Jo\v zef Stefan Institute, SI-1000 Ljubljana, Slovenia}
\affiliation{Faculty of Mathematics and Physics, University of
Ljubljana, SI-1000 Ljubljana, Slovenia}

\begin{abstract}
We develop a procedure which systematically generates all conserved operators in the disordered models of interacting 
fermions. Among these operators,  we  identify and count the independent and local integrals of motion (LIOM) which represent the hallmark of the many-body 
localization (MBL).  The method  is tested first on the prototype disordered chain  of interacting spinless fermions.  As expected for full MBL,  we find for large enough disorder
$N_M=2^M-1$ independent and quasi-local LIOM  with support on $M$ consecutive 
sites.  On the other hand, the study of the disordered  Hubbard chain reveals  that $ 3^M-1< N_M \lesssim 4^M/2$ which is less than required for full MBL
but much more than in the case of spinless fermions.

\end{abstract}

\maketitle

{\em Introduction--} The many--body localization  (MBL) \cite{Basko06,Oganesyan07} has recently attracted significant interest  as one of the most unusual phenomena in the  many-body physics. Intensive numerical studies have identified the main hallmarks of the MBL:
 vanishing of dc transport   \cite{berkelbach10,barisic10,agarwal15,lev15,steinigeweg15,barisic16,kozarzewski16,prelovsek217},
 absence of thermalization \cite{pal10, Serbyn2013, lev14,schreiber15,serbyn15,khemani15,luitz16,gornyi05,altman15,deluca13,gramsch12,deluca14,huse13,Rahul15,rademaker16,chandran15,ros15,eisert15,serbyn141,Rahul15,zakrzewski16}  and  logarithmic growth of the entanglement entropy \cite{Znidaric08,bardarson12,kjall14,Serbyn13,serbyn15,luitz16}. Very recently, it has also been found that MBL prevents a driven system from heating   \cite{kozarzewski16,abanin2017,abanin2015,Ponte2015,Bordia2017, Choi2017,Zhang2017}.

It has been shown mostly on the example of standard model of MBL, i.e. the disordered  chain of spinless fermions \cite{Serbyn2013,huse14,Rahul15,chandran15,ros15,rademaker16,ranjan15,imbrie2017}, that remarkable properties of the localized regime can be 
summarized and explained via  the existence of a macroscopic number of local integrals of motion  (LIOM). It is convenient to represent
LIOM in terms of  l--bits \cite{Serbyn2013, huse14}, which are quasi-local and mutually  commuting operators with a binary spectrum. 
The set of $l$--bits is complete in that all many-body eigenstates can be uniquely labeled with their eigenvalues \cite{imbrie2017}.  
While $l$--bits  appear to be an attractive and well established concept, the actual 
(numerical) construction of an orthogonal set of LIOM  or $l$-bits and their analysis has proven to be quite 
challenging even within the standard model of spinless fermions
\cite{chandran15,rademaker16,obrien16,inglis16}. 

The existence of LIOM, their number and presumable properties are even less established in other disordered models, in particular 
in the one-dimensional (1D) disordered Hubbard model \cite{mondaini15,prelovsek216,bonca2017,parameswaran2017}, which is actually realized and experimentally 
studied in the cold-fermion systems  \cite{schreiber15,choi16,bordia16,bordia2017_1}. It has been recently realized that in the  Hubbard chain with a potential
disorder only (as realized in the cold--atom experiments so far) the full MBL does not arise even at very large disorder \cite{prelovsek216},  
since the spin sector remains delocalized. 

It is important to clearly define the meaning of  full MBL. The easiest way is to consider first a small closed systems with $M$ sites and with the Hamiltonian 
$H=\sum_m  E_m | m \rangle \langle m| $. Since this is a finite system, all correlation functions show nonergodic behavior and all projections $|m \rangle \langle m |$  are conserved and local
(excluding the identity $\sum_m |m \rangle \langle m |$).
Then one may consider the same system but with $L \rightarrow \infty $ sites, and its finite subsystem consisting of $M$ consecutive lattice sites. Full MBL means that all local correlation functions, 
which are defined within the subsystem, are still nonergodic as if the subsystem was disconnected from the ramaining $L-M$ sites.  Therefore, in the thermodynamic limit there must be as many relevant LIOM as in a closed system of $M$ sites. In the case of spinless fermions,  there should be $N_M = 2^M-1$ orthogonal LIOM with the support on $M$ consecutive lattice sites. In the Hubbard chain, full MBL  would require $N_M = 4^M-1$ independent LIOM, while $N_M < 4^M-1$, although still macroscopic, would indicate partial MBL behavior.

In this Letter  we generalize  a numerical approach, previously designed to study local and quasi-local conserved quantities in integrable systems \cite{mm1,mm2}.
This approach allows to generate all independent LIOM in arbitrary disordered tight-binding model.
In particular, we count the LIOM and  check their locality.  In a model of spinless fermions we find  for large enough disorder $N_M= 2^M-1$  independent (orthogonal) LIOM, 
compatible with the full MBL. As a central result of this work we show that in the Hubbard model   at large disorder there are  $N_M \sim 4^M$ and $4^M/2$ LIOM, respectively, 
with and without random magnetic field.  Consequently, the disordered Hubbard model contains more LIOM  than spinless fermions, even  if the disorder is restricted  to the charge sector. 
Still, in the latter case the number of LIOM is too small to allow for full MBL.

{\it Method for constructing LIOM  --}  The general idea is similar to the approach  which has previously been used for identification of new integrals of motion 
 in the Heisenberg model \cite{mm1,mm2}.    We study a 1D disordered tight-binding Hamiltonian on $L$ sites. We pick up $M$ consecutive sites labeled with lattice index $i=1,2,...,M$ and  consider local Hermitian operators supported on these sites   $A= A_1  \otimes ... \otimes A_M  \otimes \openone_{M+1}\otimes..... \openone_{L} $, where all single--site operators 
 following $A_M$ are the identity operators.  The complete set of LIOM can be  constructed from the stiffnesses 
 $\langle \bar{A} \bar{B} \rangle=  \mathrm{Tr}(  \bar{A} \bar{B})/\mathrm{Tr}(\openone) $  where $\bar{A}$ and $\bar{B}$ are time averaged operators
 \begin{eqnarray}
\bar{A}&=&\lim_{\tau \rightarrow \infty} \frac{1}{\tau}\int_0^{\tau} {\mathrm d}t' \exp(iHt') A \exp(-iHt')  
\nonumber \\
&=& \sum_{m} |m\rangle  \langle m | A |m \rangle  \langle m | . \label{ta}
\end{eqnarray}
The latter equality holds true provided that there  are no degenerate eigenstates in a disordered system.
Our approach is based on the following facts:  the thermal--averaging  is carried out for infinite temperature  
when $\langle ... \rangle$ is also the (Hilbert-Schmidt) scalar product in the space of local Hermitian operators;  
time averaging is an orthogonal projection, i.e., $\langle \bar{A} \bar{B} \rangle=\langle A \bar{B} \rangle=\langle \bar{A} B \rangle$;
time averaged operators  are conserved by construction, i.e. $[H,\bar{A}]=0$ but they may still be nonlocal or may vanish in the thermodynamic limit, $M/L \rightarrow 0$ . 
However, a nonzero stiffness 
$D_{AB}=\langle \bar{A} \bar{B} \rangle =\langle \bar{A} B \rangle$ obtained in the thermodynamic limit for strictly local operator $B$ means that $\bar{A}$  is local or quasi-local and is the therefore a LIOM.  An important  step  (apparently missing in previous studies)  is to ensure the independence of LIOM. 

In order to find all independent LIOM, we first define a complete orthonormal basis 
$\langle O_{a} O_{b} \rangle =\delta_{a,b}$,  in the space  of local traceless operators $\{A\}$ with the support on $M$ sites.
Typically, within a fermionic model we construct $O_a$ in terms of identity 
($ \openone_{i}$), creation ($c^{\dagger}_{i}$ or $c^{\dagger}_{i,\sigma}$), annihilation ($c_{i}$ or $c_{i,\sigma}$) 
and the particle--number ($n_{i}$ or $n_{i,\sigma}$)  operators. Then, we diagonalize Hamiltonian and construct the time--averaged 
operators, $\bar{O}_a$,  as defined in Eq. (\ref{ta}).  The last step is to diagonalize the  matrix of stiffnesses
\begin{equation}
\sum_{a,b} U^T_{\alpha a} \langle \bar{O}_{a} \bar{O}_{b} \rangle  U_{b \beta}=\lambda_\alpha \delta_{\alpha \beta}, \label{eig}
\end{equation}
and  to rotate the original basis 
\begin{equation}
Q_{\alpha}=\sum_{a} U_{a \alpha} O_a, \quad \quad \bar{Q}_{\alpha}=\sum_{a} U_{a \alpha} \bar{O}_a.
\end{equation} 
Since $\langle \bar{Q}_{\alpha} \bar{Q}_{\beta} \rangle=\langle \bar{Q}_{\alpha} Q_{\beta} \rangle=\lambda_\alpha \delta_{\alpha \beta}$, 
one ends up with strictly conserved and orthogonal (hence independent) operators  $\bar{Q}_{\alpha}$
which contain local parts determined by the projection  $\langle \bar{Q}_{\alpha} Q_{\alpha} \rangle = \lambda_\alpha$.  In other words, 
\begin{equation} 
\bar{Q}_{\alpha}=\lambda_{\alpha} Q_{\alpha}+Q^{\perp}_{\alpha}, \label{qq}
\end{equation}
 where  $Q^{\perp}_{\alpha}$ is the nonlocal part  of $\bar{Q}_{\alpha}$, i.e., $\langle Q^{\perp}_{\alpha} Q_{\beta} \rangle=0$ for any $Q_{\beta}$.
In the extreme case one gets $\lambda_{\alpha}=1$ and the corresponding LIOM is strictly local, i.e., $\bar{Q}_{\alpha}=Q_{\alpha}$. 
However, if  we get $0<\lambda_{\alpha}<1$ for $L \gg M$ then the corresponding LIOM, $\bar{Q}_{\alpha}$,  is quasilocal since $Q^{\perp}_{\alpha} \ne 0$.
 

The stiffnesses (i.e., the long--time dynamics) of all local operators (with maximal support on $M$ sites) are now fully determined by their projection on the complete local
basis $Q_\alpha$,  $A=\sum_{\alpha} \langle A Q_{\alpha} \rangle Q_{\alpha} $. In particular, the inequality known as the Mazur bound \cite{zotos97}
 turns in the present case into equality for the stiffnesses 
 \begin{eqnarray}
D_{AB} &=& \langle  \bar{A}\bar{B} \rangle = \sum_{\alpha}  \lambda_{\alpha}  \langle A Q_{\alpha} \rangle \langle B Q_{\alpha} \rangle \nonumber \\ 
& = & \sum_{\alpha \ne 0} \frac{\langle A \bar{Q}_{\alpha} \rangle \langle B \bar{Q}_{\alpha} \rangle} {\langle \bar{Q}_{\alpha} \bar{Q}_{\alpha} \rangle} . \label{mazur}
\end{eqnarray}
While the relation is valid for any system size $L > M$, it gets its full meaning in the limit $L \to \infty$. The hallmark of MBL is then  
nonvanishing $ D_{AB} \neq 0$,  emerging from LIOM with $\lambda_\alpha >0$ [see Eq.~(\ref{mazur})]. Full MBL requires that
all local correlations are nonergodic, which can emerge if there is sufficient number of LIOM with $\lambda_\alpha >0$. This numbers should be $2^M-1$ and $4^M-1$ for spinless and spin--$1/2$ fermions, respectively.   
It is rather clear that the dynamics of local operators is determined by the "most local" LIOM, i.e., by $\bar{Q}_{\alpha}$ with the largest $\lambda_{\alpha}$.   
The main advantage of our method is that we can find all othogonal $\bar Q_\alpha$, but even more that we can sort them out by locality,
i.e. by ordering them with respect to their $\lambda_\alpha$.  

 {\it Disordered chain of interacting spinless fermions--} We first study standard model of MBL, i.e. the 1D system of  interacting spinless fermions 
 \begin{eqnarray}
 H&=& \sum_i  \left[ -t (c^{\dagger}_{i+1} c_i+{\rm H.c.})+\varepsilon_i n_i  + V n_i n_{i+1} \right], 
 \end{eqnarray} 
 where $n_i=c^{\dagger}_{i} c_i$ and $\varepsilon_i$ are uncorrelated random potentials with a uniform distribution, $\varepsilon_i \in [-W,W]$.  We express 
 all energies in units of $t=1$. We consider only the system at half-filling  with $N=L/2$ particles and the interaction  $V=1$ (if not stated otherwise).
 
 \begin{figure}[h!]
\centering
\includegraphics[width=\columnwidth]{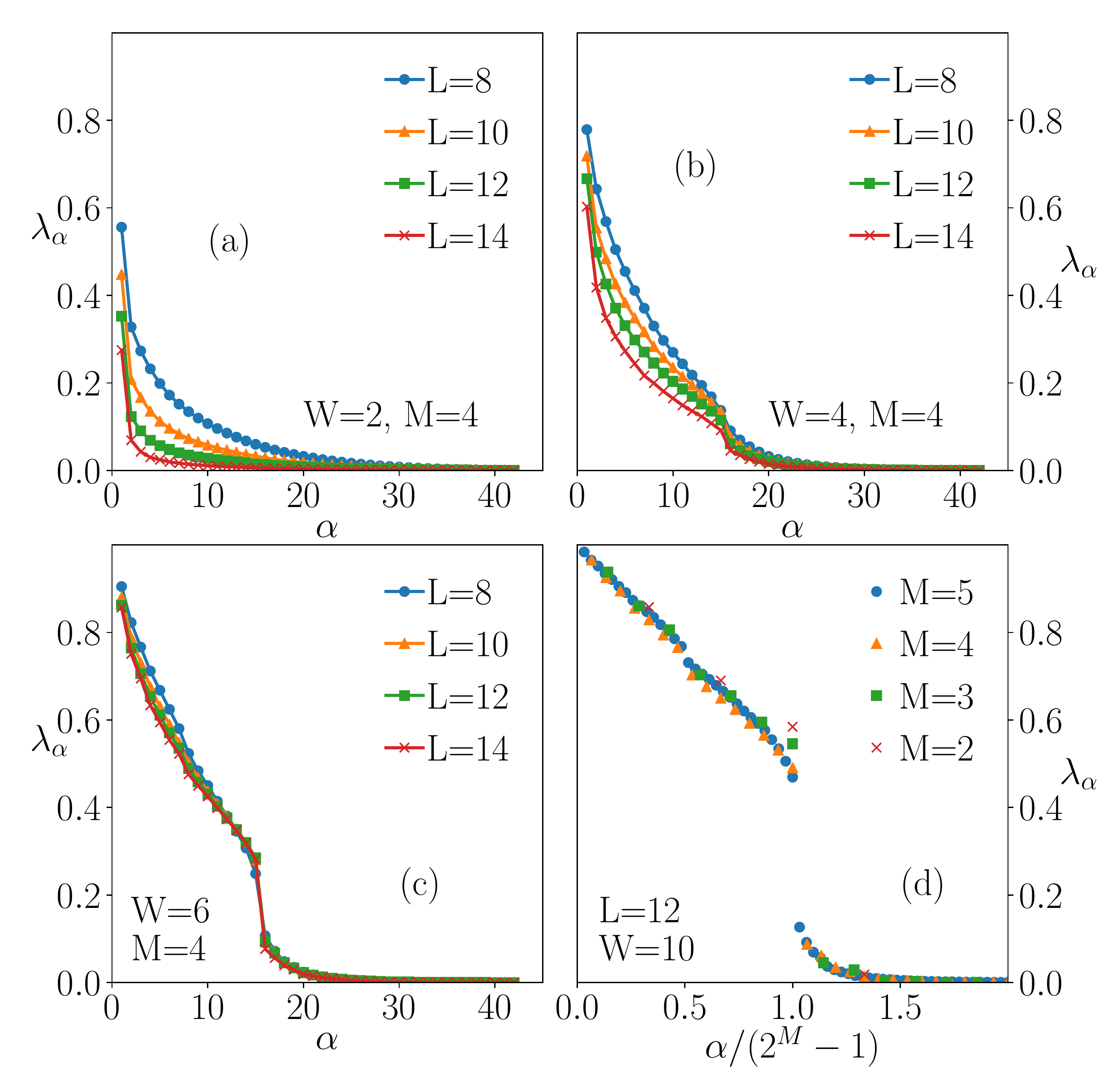}
\caption{Results for disordered chain of spinless fermions.  Eigenvalues  $\lambda_\alpha$ corresponding to local components of LIOM, see Eq. (\ref{qq}), 
averaged over disorder for various system sizes $L$, supports $M$ and different disorders $W$. 
}
\label{fig1}
\end{figure} 

The procedure described in the preceding section, has been carried out independently for 100-500 sets of random configuration $\{\varepsilon_1,...,\varepsilon_L\}$. 
For each set we obtain LIOM, $\bar{Q}_{\alpha}$, along with information on their locality which is stored in $\lambda_{\alpha}$. Since our primary aim 
is just to count the number of independent LIOM we average  $\lambda_{\alpha}$ over various realizations of disorder.   
Such averaging mostly erases information on the structure
of LIOM and this aspect  will be discussed later on.  

\begin{figure}[h!]
\centering
\includegraphics[width=\columnwidth]{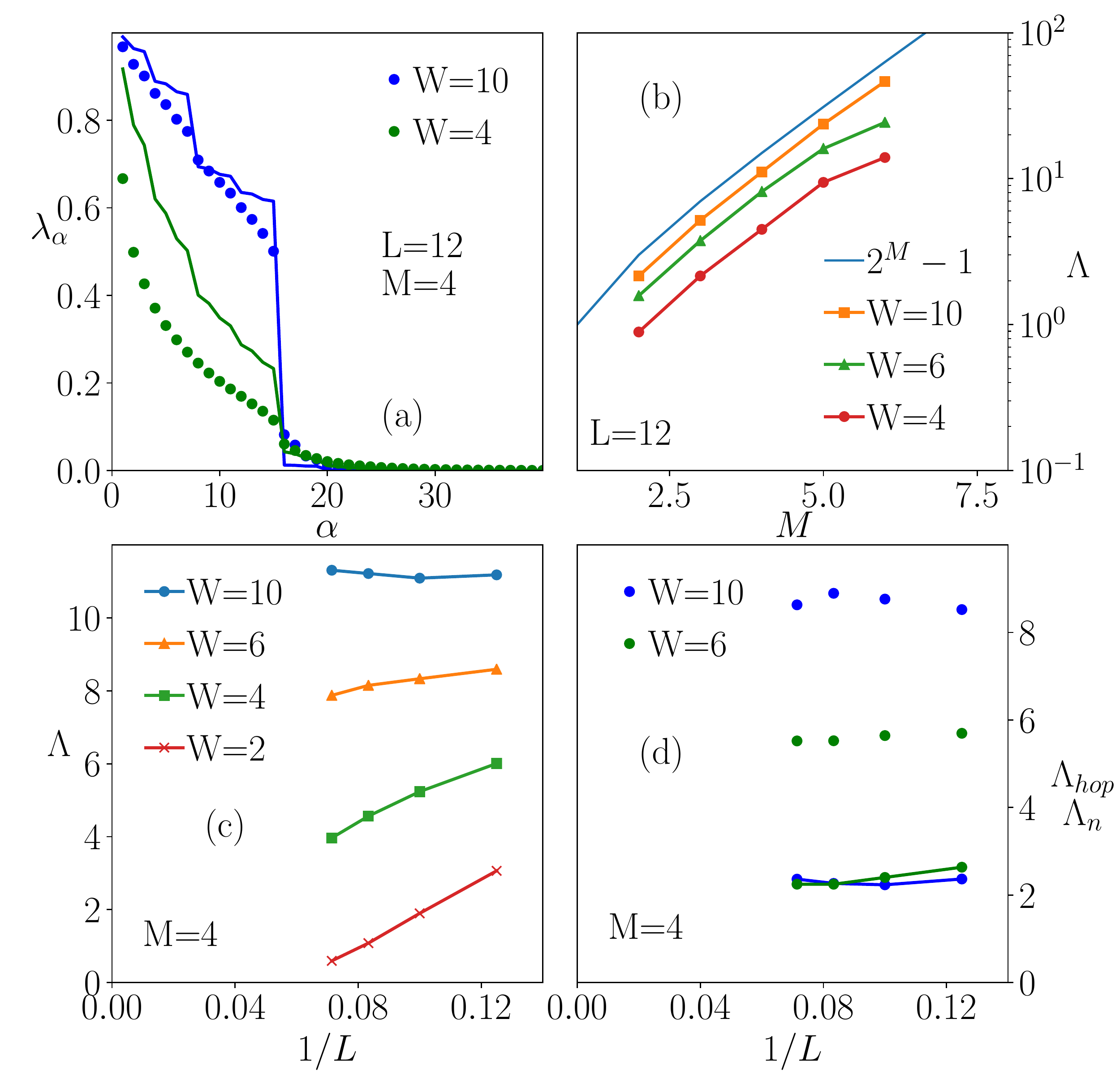}
\caption{ Results for disordered chain of spinless fermions. (a) Sorted eigenvalues $\lambda_\alpha$ (local components of LIOM)  for $V=1$  (points) and for Anderson insulator, 
$V=0$ (lines), (b) total weight of LIOM $\Lambda$ for various supports $M$ and different disorders
$W$, (c) finite--size scaling of $\Lambda$ with $1/L$ for $M=4$ and different $W$, (d) total weights vs. $1/L$ constructed from local operators 
with at least one hopping term  ($\Lambda_{hop}$ -- points with lines) or without hopping  ($\Lambda_{n}$ -- points) for $W=6,10$. } 
\label{fig2}
\end{figure} 

In the present case ($V=1$) the MBL transition is at $W_c \sim 4$ \cite{barlev15,mierzejewski16}.
Figs.~\ref{fig1}a,b,c show the averaged $\lambda_{\alpha}$ below the MBL transition ($W=2$), close to the transition   ($W=4$) 
and in the MBL regime  ($W=6$), respectively.  In the latter case (fig. \ref{fig1}c), there is a clear jump in $\lambda_{\alpha}$. 
For $M=4$, it separates the LIOM $\bar{Q}_1,...,\bar{Q}_{15}$ from other conserved
 quantities  $\bar{Q}_{16},...$, which have only marginal overlap with local operators.  
 These results are almost independent of the system size provided that $L \gg M$, hence, 
 they should be true also for $L \to \infty$.  The  jump  in $\lambda_{\alpha}$ starts to form at the MBL transition (Fig.~\ref{fig1}b) and is absent  
 in the ergodic phase (Fig.~\ref{fig1}a). In the ergodic system all $\lambda_{\alpha}$ (at fixed $M$)
also vanish with increasing $L$,  as evident from Fig.~\ref{fig1}a  and shown later on for  
 $L \rightarrow \infty$.  Fig.~\ref{fig1}d  shows that the position of the jump
 depends on the support $M$ of local operators, i.e. there are exactly $N_M=2^M-1$ nontrivial LIOM. This number is consistent with 
 full MBL  and with the presence of $M$ relevant l-bits, $\sigma_1,...,\sigma_M$,  
 which build the LIOM, e.g. $\bar{Q}_{\alpha}=\sigma_i$ or  $\bar{Q}_{\alpha}=\sigma_i\sigma_j\sigma_k$.  
In Fig.~\ref{fig2}a we explicitly show, that the number   of LIOM, $N_M$, in the MBL system is exactly the same as in the Anderson insulator ($V=0$).   
 
The number of LIOM  in the vicinity of the MBL transition is the same as deep in the MBL phase,  whereas their locality is quite different in these regimes. 
An important message coming from Eq.~(\ref{mazur}) is that  the number $N_M$ itself does not explain the actual value of nonergodic $D_{AB} \neq 0$. 
In contrast to strictly local charges, proper counting of quasi-local LIOM should be weighted by their "locality".  Therefore, we introduce the total weight  of LIOM,
$\Lambda=\sum_{\alpha} \lambda_{\alpha}$. In case of strictly local LIOM ($\lambda_{\alpha}=1$)  we would get $\Lambda = N_M$.  Fig.~\ref{fig2}b shows
that $\Lambda$ indeed increases with $M$ as $\Lambda \propto N_M $ but as well reveals an overall decrease when approaching the MBL transition at $W \sim W_c$.  

Figure~\ref{fig2}c  shows finite--size  scaling of $\Lambda$ vs. $1/L$ for fixed $M=4$. 
Results indicate  that $\Lambda$ vanishes in the limit $L \to \infty$ at the MBL transition $W = W_c \simeq 4$, while remaining finite for $W>W_c$. 
It means that all time--averaged operators become nonlocal and consequently  all stiffnesses vanish, i.e. the  system becomes ergodic. 

Finally, let us comment  the structure of LIOM as it is possible to deduce from the disorder--averaged quantities.
 For this reason we restrict the space of local operators  and separately study the cases when  the basis operators $O_{a}$  contain only  
 $\openone_i$ or  $n_i -1/2$ and when  each $O_{a}$ contains at least one hopping term, $c^{\dagger}_i c_{j\ne i}$.  
 In this way we obtain the total weights $\Lambda_n$ and $\Lambda_{\rm hop}$, respectively. It is easy to show  
that $\Lambda=\Lambda_n + \Lambda_{\rm hop}$, hence $\Lambda_n$ and $\Lambda_{\rm hop}$ represent contributions to $\Lambda$ coming from both types of operators. Changing the disorder strength within the MBL regime one modifies mostly  $\Lambda_n$, which becomes the dominating contribution for large $W$ (see  Fig.~\ref{fig2}d).  Therefore, 
deep in the MBL regime  l--bits can be well approximated by $\bar{n}_i-1/2$.

 {\it Disordered Hubbard chain--}  Next, we turn to the main focus of this work and study the disordered Hubbard chain,
\begin{eqnarray}
H&=& -t \sum_{i,\sigma}   (c^{\dagger}_{i+1,\sigma} c_{i,\sigma}+{\rm H.c.})+\sum_{i,\sigma} \varepsilon_i  n_{i, \sigma}\nonumber \\ 
   && + U \sum_i  n_{i,\uparrow} n_{i,\downarrow}, \label{hh}
 \end{eqnarray} 
 where $n_{i ,\sigma}=c^{\dagger}_{i,\sigma} c_{i,\sigma}$ and disorder enters (so far) only in the (charge) potential. 
Since the dimension of the Hilbert space grows as $4^L$,  it is more demanding to satisfy the requirement $L \gg M$. Hence, we do not attempt 
a proper finite-size scaling, but with respect to the MBL transition we rely on (few) previous studies of  the model \cite{mondaini15,prelovsek216}. However,
we can still accomplish our main goal of counting the LIOM.   For this sake, we restrict our studies mostly  to strong disorder, $W=15$, where finite--size
effects are supposed to be negligible (see Fig.~\ref{fig2}c).  Our numerical calculations have been carried out for $L=8$. We study a half--filled system with $N=4$ fermions for each spin projection, i.e.  we consider the sector $S^z_{tot}=0$. Results for $U=2, 4$ or $8$ (not shown) are very similar to the data presented in this work for $U=1$.

\begin{figure}[h!]
\centering
\includegraphics[width=\columnwidth]{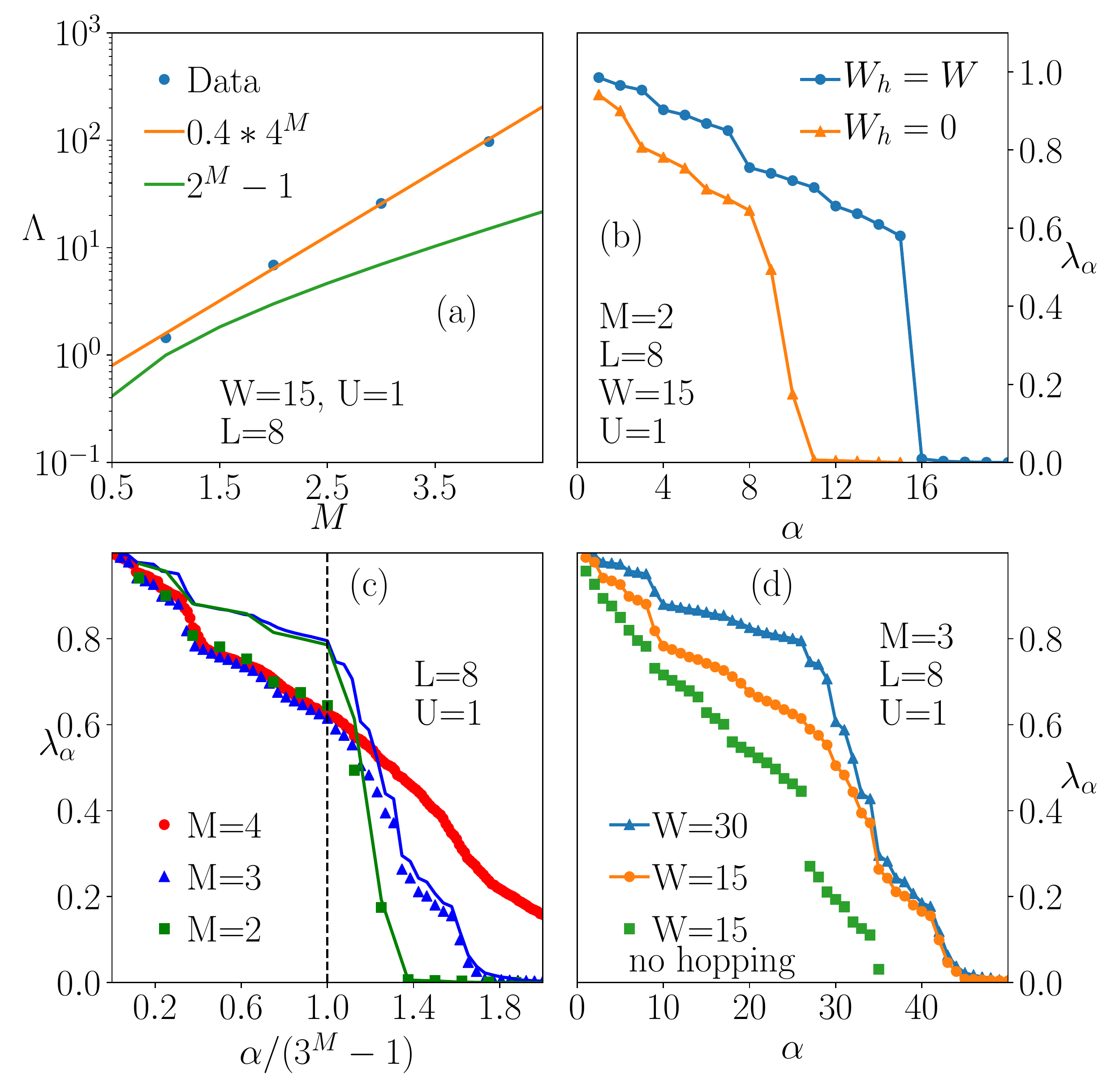}
\caption{Results for disordered Hubbard chain. (a) total weight $\Lambda$ of LIOM (points) for  various supports $M$ together with two different estimates; 
(b),(c),(d): sorted averaged eigenvalues $\lambda_\alpha$ of LIOM.  (b) results for LIOM constructed for Hubbard chain without and with 
random magnetic field, respectively. Points and lines in (c) show results for $W=15$ and $W=30$, respectively. (d) LIOM constructed from local operators with and without hopping terms, respectively. 
 }
 \label{fig3}
\end{figure} 

Fig.~\ref{fig3}a shows the total weight $\Lambda$ of LIOM as a function of $M$. It is clear that $\Lambda$ by far exceeds the number of LIOM in the system
of spinless fermions and grows much faster than $2^M$. The best fit, $\Lambda \sim 0.4 \times 4^M$,
might suggest that the system comes close to full MBL. Nevertheless, localization of all degrees of freedom would require, in analogy to spinless fermions,  $N_M = 4^M-1$ and 
$\Lambda \simeq 4^M-1$ (in the large-disorder regime). 
In order to show the existence of such a case,  we study also a modified Hubbard chain (\ref{hh})
 with a random magnetic field $H \rightarrow H + \sum_i h_i (n_{i,\uparrow}-n_{i,\downarrow})  $, where $h_i \in [-W_h,W_h]$ and $W_h=W$.  
 Indeed,  the latter case leads to full MBL with exactly $N_M=4^M-1$ LIOM, as clearly  shown in Fig.~\ref{fig3}b for $M=2$. However, 
 without random magnetic field ($W_h=0$), we get $N_M \sim 4^M/2$ which excludes the full MBL \cite{prelovsek216,su21,su22,su23}.  
 In the case of $W_h=W$,  LIOM have significant projections on operators which are either even or odd under 
 the spin--flip transformation, whereas only even LIOM exist at $W_h=0$.   
 
 It is harder to count the relevant LIOM directly from their $\lambda_{\alpha}$.  
The largest components $\lambda_{1}, ...., \lambda_{3^M-1}$   form a plateau shown in Fig.~\ref{fig3}c.
Within this plateau, we observe a pretty convincing scaling of $\lambda_{\alpha}$ similar to that in Fig.\ref{fig1}d.
 Beyond this plateau, there are other LIOM which significantly contribute to the dynamics of local operators. However,  already the number of  
 LIOM within the plateau, $\tilde N_M \sim 3^M-1$,  exceeds the result for spinless fermions,  $2^M-1$.   
 In order to explain the origin of LIOM within this plateau, we have carried out calculations also for the restricted space of local basis
operators $O_a$, 
 excluding operators which contain  hopping terms.  Results in Fig.~ \ref{fig3}d  show that within this restricted space, there is again
 a clear jump in $\lambda_{\alpha}$ exactly at $\alpha=3^M-1$. 
We expect that operators from this restricted space  give the dominating contribution to the most local LIOM in the Hubbard chain. 
Then, it is also easy the explain the $3^M$ scaling. Namely, the LIOM within the plateau arise from  the time--averaging of basis operators, $O_a$, built
out of three SU(2) invariant, single--site operators: $\openone_i$,  $\sqrt{2}(n_{i,\uparrow}+n_{i,\downarrow}-1)$ and $(2n_{i,\uparrow}-1)(2n_{i,\downarrow}-1)$.  

{\it Conclusions--} We have introduced and implemented a general method for identifying independent LIOM in disordered tight-binding models. 
Studying a system of spinless fermions, we have shown that this approach  allows one to locate the MBL transitions as  a regime where {\it all} LIOM vanish.  
Most importantly, it allows to count LIOM which, by construction, are orthogonal. We have shown that the stiffnesses of 
 operators defined within a subsystem of $M$ sites are governed by $N_M=2^M-1$ independent LIOM.  $N_m$ is thus the same as the maximal number of orthogonal conserved operators (projections on the eigenstates of Hamiltonian) in a closed system of $M$ sites, supporting the full MBL. 
$N_M$ is independent of disorder $W$, but operators are quasi-local only for large enough disorder $W > W_c$ and do not have property of LIOM for $W <W_c$. 
 
Our main results concern the number of LIOM in the disordered Hubbard chain. We have explained the origin of the $3^M-1$ most 
local LIOM which are relevant for the long--time correlations (stiffnesses) of observables on
$M$ sites. The total weight of LIOM (at large potential disorder) seems to grow as $\Lambda
\sim  4^M/ 2$. Therefore, the number of LIOM in the Hubbard chain is much larger than in the system of 
spinless fermions. However, unless one applies a (large enough) random magnetic field, the number of LIOM is  still 
smaller than required for full MBL. 

 \acknowledgments  
  This work is supported by the National Science Centre, Poland via project 2016/23/B/ST3/00647 (MM and MK) and P.P. acknowledges the support by the program P1-0044 of the
Slovenian Research Agency.
    
\bibliography{references.bib}

\end{document}